\documentclass[twocolumn, trackchanges]{aastex62}
\pdfoutput=1 
\usepackage{amsmath,amstext}
\usepackage[T1]{fontenc}
\usepackage{apjfonts} 
\usepackage[figure,figure*]{hypcap}

\shorttitle{MHD and CNNs}
\shortauthors{Peek \& Burkhart}

\begin{document}

\title{Do Androids Dream of Magnetic Fields? Using Neural Networks to Interpret the Turbulent Interstellar Medium}

\author[0000-0003-4797-7030]{J. E. G. Peek}
\affil{Space Telescope Science Institute, 
3700 San Martin Drive, 
Baltimore, MD, 21218}

\affil{Department of Physics \& Astronomy, 
Johns Hopkins University,
3400 N. Charles Street, 
Baltimore, MD 21218}

\author[0000-0001-5817-5944]{Blakesley Burkhart}
\affil{Center for Computational Astrophysics, 
Flatiron Institute, 
162 Fifth Avenue, 
New York, NY 10010}
\affil{Department of Physics and Astronomy, 
Rutgers University, 
136 Frelinghuysen Rd, 
Piscataway, NJ 08854}

\begin{abstract}
The interstellar medium (ISM) of galaxies is composed of a turbulent magnetized plasma.  In order to quantitatively measure relevant turbulent parameters of the ISM, a wide variety of statistical techniques and metrics have been developed that are often tested using numerical simulations and analytic formalism.  These metrics are typically based on the Fourier power spectrum, which does not capture the Fourier phase information that carries the morphological characteristics of images. In this work we use density slices of magnetohydrodyanmic  turbulence simulations to demonstrate that a modern tool, convolutional neural networks, can capture significant information encoded in the Fourier phases.  We train the neural network to distinguish between two simulations with different levels of magnetization.  We find that, even given a tiny slice of simulation data, a relatively simple network can distinguish sub-Alfv\'enic (strong magnetic field) and super-Alfv\'enic (weak magnetic field) turbulence $>$98\% of the time, even when all spectral amplitude information is stripped from the images. In order to better understand how the neural network is picking out differences betweem the two classes of simulations we apply a neural network analysis method called ``saliency maps''.  The saliency map analysis shows that sharp ridge-like features are a distinguishing morphological characteristic in such simulations. Our analysis provides a way forward for deeper understanding of the relationship between magnetohydrodyanmic turbulence and gas morphology and motivates further applications of neural networks for studies of turbulence. We make publicly available all data and software needed to reproduce our results. 
\end{abstract}

\section{Introduction}

Magnetohydrodyanmic (MHD) turbulence is now part of the established paradigm of the interstellar medium (ISM) of galaxies and can influence the behavior of gas and dust over scales ranging from beyond a kiloparsec to below AU \citep{Armstrong95,ElmegreenScalo,Lazarian07rev,burkhartphd}.   
MHD turbulence is understood to be important for fundamental astrophysical processes including star formation, galactic pressure support, the transport of heat and metals, formation of molecular hydrogen, the acceleration and diffusion of cosmic rays and the structure of magnetic field lines\citep{lazarianvishniac99,maclow04,lazarian06,Yan09,burkhart10,Burkhart12,federrath12,bialy2017ApJ...843...92B,Pingel2018ApJ...856..136P}.

Despite the importance of MHD turbulence for galactic processes, it is difficult to devise metrics which can accurately characterize plasma fluid  properties from astronomical observations.  MHD turbulence is notoriously difficult to study even in a controlled laboratory setting \citep{Nornberg2006,Bayliss2007} and even more challenging to quantify and understand when dealing with line of sight (LOS) effects, radiative transfer, and telescope beam smearing \citep{Hill2008,burkhart13a,Offner14a,2019arXiv190410484K}.
In light of these challenges, direct numerical simulations of turbulence have been critical for our understanding of the physical conditions and statistical properties of MHD turbulence in astrophysical environments \citep{Cho:2002gz,maclow04,Ballesteros-Paredes07a,Mckee_Ostriker2007,Kowal07,Federrath08a,Burkhart09a,Collins12a}.
Numerical MHD turbulence simulations generally can resemble observations in terms of the scaling of the power spectrum and the overall density structure (e.g. filaments). However numerical simulations of turbulence lack the spatial dynamic range of the real ISM and therefore cannot reach the observed Reynolds numbers of nature. The Reynolds number ($R_e^{1/2}$) is the ratio of the large eddy turnover rate to the viscous dissipation rate. Therefore large $R_e$ correspond to negligible viscous dissipation of large eddies over the eddy turn over time and simulations are therefore too viscous, usually due to numerical viscosity. 
However, the statistical properties of the ISM, including the density histogram (or PDF) \citep{vazquezsemadeni97,Federrath09a,Burkhart09a,Kainulainen09a,Burkhart12, burkhart15,Chen2018ApJ...859..162C,Burkhart2017ApJ...834L...1B}, velocity/density power spectrum \citep{Stanimirovic99a,lp00,Stanimirovic04a,LP08,burkhart10,Bur13,chepurnov15,Pingel2018ApJ...856..136P}, three point functions/phase analysis  \citep{Burkhart09a,Burkhartlaz16,Portillo2018,2019arXiv190410484K} and principle component analysis \citep{Heyer04a,Hey12,Yuen2018ApJ...865...54Y}, are well represented by MHD turbulence simulations.  
Statistical studies therefore may be the current best method for understanding the properties of turbulence and for connecting observations and numerical simulations.

The most common statistical tool of turbulence studies for nearly a century has been the spatial or temporal Fourier power spectrum. 
This is because the power spectrum enables the examination of the turbulence energy cascade as a function scale or frequency and can reveal the sources (injection scale) and sinks (dissipation scale) of energy and the self-similar behavior of the inertial range scaling. Furthermore, the power spectrum of ISM density and LOS Doppler shifted velocity can inform on the spatial and kinematic scaling of turbulence and sonic Mach
number  \citep{LP04,LP06, Kowal07, goodman09,Heyer09a,burkhart10,Collins12a, burkhart14,chepurnov15,2019arXiv190410484K}.

The power spectrum is defined as:

\begin{equation}
P(k)=\sum_{k}\tilde{F}(k)\cdot\tilde{F}^{*}(k)
\label{eq:ps}
\end{equation}

where $k$ is the wavenumber and $\tilde{F}(k)$ is the Fourier transform of the field under study, e.g. density, velocity, magnetic energy etc.

Equation \ref{eq:ps} demonstrates a critical limitation of the
Fourier power spectrum: it contains only the
 Fourier \textit{amplitudes} and neglects the Fourier phases. This is problematic for studies of MHD
turbulence because interactions among MHD waves can
produce correlations in Fourier phases which are completely
missed by the power spectrum \citep{Burkhartlaz16}. The structure imprinted on the phase information will be entirely lost to a power spectral analysis. Furthermore, the degree of phase coherence or
randomness in MHD turbulence is important for MHD  wave-wave interactions and particle transport.  In order to study both phase and amplitude information a number of statistics beyond the power spectrum have been proposed for ISM studies, including the 3 point correlation function or bispectrum \citep{Burkhart09a,Portillo2018} and the phase coherence\citep{Burkhartlaz16}. 
Additional statistical tools and methodologies for studies of turbulence have been developed in the last several decades and include course graining \citep{Aluie_2017,PhysRevX.8.011023,PhysRevLett.122.135101},
higher-order moments \citep{Kowal07,Burkhart09a,burkhart10,2011Natur.478..214G}, wavelets \citep{2010ApJ...720..742K,farge_schneider_2015,doi:10.1063/1.5062853}, topological
techniques \citep{Kowal07,2008ApJ...688.1021C, 2012ApJ...749..145B}, clump and hierarchical structure finders \citep{Rosolowsky08a, goodman09,burkhart13}, Tsallis
distributions \citep{Esquivel11,tofflemire11}, and structure functions as tests of intermittency and
anisotropy \citep{1994PhRvL..72..336S,CL03,esquivel05,KowL10,2014ApJ...790..130B,2015ApJ...804..119V,2018ApJ...857L..19H,PhysRevLett.122.135101}.
While these tools have been successful at obtaining some of the turbulence parameters such as the driving scale and sonic Mach number, they are less adept at obtaining magnetic field information that might be encoded in the phases and are somewhat difficult to interpret physically.

A generic problem in studying the phase content of images is that there is no single metric that encompasses all the information, and thus feature vectors  must be constructed ``by hand'', informed either by our best theoretical understanding of the underlying processes or by observations that clue us in to important morphological properties \citep[e.g.][]{2014ApJ...789...82C} or are otherwise difficult to glean physical meaning from, such as the bispectrum \citep{burkhart10}.
Nevertheless, the phase information is crucial for reproducing images with content understandable by humans. For example, \cite{1981IEEEP..69..529O} showed that preserving the phase in an image but scrambling or distorting the power generated images that are comprehensible to the human eye, but preserving the power and and distorting the phase led to incomprehensible images. 
This result gives us the intuition that a tool or statistic that can distinguish between visually distinct structures may be able to tap into the information locked up in the Fourier phases.

Tools that mimic the human capacity to distinguish between physically and semantically different scenes have long been the goal of Computer Vision and Machine Vision. These disciplines have tended to build up semantic information from low-level, custom built feature vectors from segmented images into ever more complex scene analysis tools \citep[e.g.][]{Sonka08}. The rise of neural networks, deep learning, and, in particular convolutional neural networks (CNNs) has completely revolutionized our ability to extract semantic information from images, and upended the computer vision paradigm \citep[see e.g.][for a broad overview]{LeCun:2015dt}. CNNs allow for hierarchical feature extraction without any custom feature vectors, and as such they can be used for astronomical images as easily as other images. Indeed, more than 40 papers have been published using CNNs in the astronomical literature in 2018, twice the number in all previous years combined. While many of these articles are exploratory in nature, a fair number report increased speed and/or increased accuracy over previous methods.

A common complaint among physical scientists is that while neural networks, and machine learning systems as a whole, can accurately propagate labels and perform regressions, they represent a ``black box''. These systems often have millions of fit parameters, and thus understanding what the system has ``learned'' is far from trivial. In recent years, significant progress has been made in extracting the knowledge built up by a neural network by applying new analysis tools to understand the network's intuition (see e.g. \url{distill.pub}). Understanding the way neural networks operate in different physical contexts may allow us to gain insight into how information is encoded in astronomical images, and thereby build more powerful physical theories.

In this work we demonstrate the power of CNNs for studies of ISM turbulence in the absence of power spectral information, i.e. just using the pure phase information. We take a number of steps to remove information that is known to be useful in the studies of magnetized turbulence to demonstrate how CNNs can harness currently unknown aspects of the information contained in images to measure the physics of turbulence. We will show that we can detect changes in the Alfv\'en Mach number without velocity information, histogram information, the power spectrum, or stretched features along the background magnetic field.

This paper is organized as follows:
In \S 2 we describe the simulation data we use in the analysis. In \S 3 we describe CNNs and the network architecture and methods we use. In \S 4 we report our quantitative and qualitative results. We discuss and conclude in \S 5.

\section{Simulation Data}

We use the database of 3D numerical simulations of
isothermal compressible (MHD) turbulence with resolution
$512^3$ presented in a number of past works \citep{Kowal07a,Burkhart09a,burkhart13a}
This database of simulations is part of the Catalog for Astrophysical Turbulence Simulations (CATS, Burkhart et al. 2019, in prep)\footnote{More information on the CATS project can be found at www.mhdturbulence.com}.
We refer to these works for the
details of the numerical set-up and here provide a short
overview.
 The code is a third-order accurate
ENO scheme \citep{cho03} which solves the ideal MHD equations
in a periodic box with purely solenoidally driving and uses an isothermal equation of state.
We vary the input values for the sonic Mach number
($M_s = v/c_s$, where $v$ is the flow velocity and $c_s$ is the
sound speed) and Alfv\'enic Mach number ($M_A = v/v_A$,
where $v_A$ is the Alfv\'en speed).
The magnetic field consists of the uniform background
field and a turbulent field, i.e: $B = B_{ext} + b$. Initially $b
= 0$ and the initial density field is uniform.

We test the utility of neural networks for finding magnetic field structure in density using snapshots of each of two simulations at a resolution of $512^3$.
The simulations have a fixed sonic Mach number of 
$M_s$=7. There are two different magnetic field values
used in this investigation: $M_A\approx 0.7$ (sub-Alfv\'enic) and
$M_A\approx 2.0$ (super-Alfv\'enic), otherwise the simulations examined have identical parameters. We use four snapshots from each simulation at 5, 5.5 6, and 7 eddy turnover times.

The sonic and Alfv\'enic Mach numbers are important control parameters for simulations of MHD turbulence. They are also related to the plasma $\beta$, i.e. the ratio of the thermal pressure ($P_{\rm thermal}$) to magnetic pressure ($P_{\rm mag}$), which is defined as $\beta=\frac{P_{\rm thermal}}{P_{\rm mag}}$. The Plasma $\beta$ can be defined as $\beta=\frac{2M_A^2}{M_s^2}$.
The sonic Mach number can be measured directly from the LOS velocity dispersion of spectral lines and measurements of temperature in the ISM.  However because magnetic fields are difficult to directly detect \citep{Crutcher:2012hw,Soler:2013dh,Clark:2012bq,Lazarian2018ApJ...865...46L}, the Alfv\'enic Mach number is more elusive. The simulation data used are available here: \dataset[10.7910/DVN/UKOPYP]{https://doi.org/10.7910/DVN/UKOPYP}.

\section{Method: Convolutional Neural Networks (CNN)}

Our goal is to build a system that can distinguish between simulations sub- and super- Alfv\'enic turbulence, and thereby learn what information exists in the image plane about the turbulence.
To do this we use a CNN deployed with the Keras and Tensorflow frameworks \citep{tensorflow2015-whitepaper,chollet2015keras}.

CNNs (also called \emph{convnets}) are a subclass of neural networks that have a specific limited connectivity, originally patterned after the structure of the neural structure of the visual processing structures in brains \citep{DHHubel:1962ep}. They were first made popular in the astronomical literature by \cite{Dieleman:2015fk} who used a CNN to reproduce visual classifications of galaxy morphology. 

A standard neural network is composed of \textit{neurons}: a neuron takes in some inputs and provides an output
\begin{equation}
y = f\left(\sum_{i=1}^d x_i \centerdot w_i \right),
\end{equation}
where $x_i$ is the input vector, $w_i$ are a set of trainable weights, and $f$ is some non-linear, typically monotonically increasing, activation function \citep[e.g.][]{LeCun:2015dt}. In this way a network can be built up connecting the outputs of many neurons on some layer $j$ to many neurons on the next layer $j+1$. A ``fully connected layer'' is a a set of weights between layers of neurons that connects all neurons on level $j$ to level $j+1$. A network will typically narrow, until the number of final neurons is equal to the number of classes the network is being designed to distinguish. The process of training a network consists of providing labeled data to the network and measuring the resultant class. If the class is incorrect the weights are adjusted through back-propagation of error. In most modern training scenarios a small subset of the training data is presented to the network at a time, and the weights are adjusted in a process called \textit{stochastic gradient descent}.

A CNN departs from this standard structure by the introduction of two extra layer types, \textit{convolutional layers} and \textit{pooling layers.}\footnote{See \url{https://cs231n.github.io/convolutional-networks/} for a discussion} If we construct our input as an $M \times M$ image, a first convolutional layer would map each $N \times N$ sub-patch of the image to a neuron in the next layer below with a grid of weights $w_{ij}$. This in effect generates a convolution of the image above. In practice many such convolutions would be applied to the layer, represented by $w_{ij,k}$, thus mapping to a $M \times M \times K$ grid of neurons. A pooling layer decreases the resolution of the image by only passing the maximum activation of the neurons within a patch to the next layer. Thus, by alternating pooling and convolutional layers, the grid of neurons becomes smaller in the image plane and deeper in the convolutional dimension until fully connected layers are used at the end of the network. A more complete description of CNNs appropriate for an astronomical audience can be found in \S 5 of \cite{Dieleman:2015fk}.

\begin{figure}[t!]
\begin{center}
\includegraphics[width=3.5in]{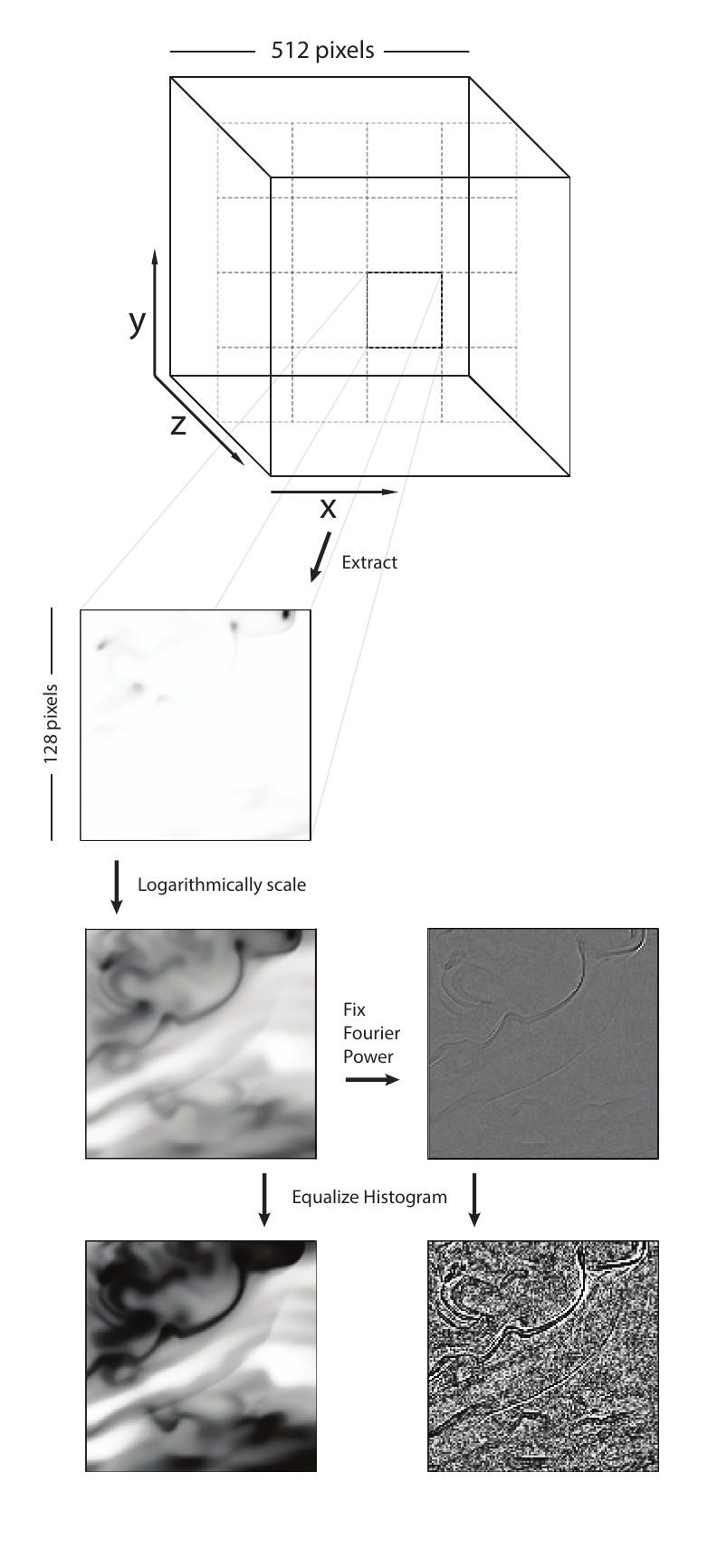}
\caption{Preprocessing steps for the data. Each slice of each simualtion data cube is split into 16 128 x 128 grid point regions and extracted as an image. We then take the logarithm of the image pixels. These images are then split into two groups, one is left as is, the other is then run through the "Fixed Fourier Power" procedure. Then both sets of images are histogram equalized.}
\end{center}
\label{prep}
\end{figure}

CNNs take input images that typically go through a number of preprocessing steps. Our preprocessing steps are designed both to make the network effective, but also to strip out as much information as possible not relevant to our study of phase information (see Figure \ref{prep}). To make images to start with we extract 128 x 128 slices from the 512 x 512 x 512 grid point density cubes perpendicular to the magnetic field. 
Turbulent eddies are stretched along the magnetic axis \citep{Goldreich95a,lazarianvishniac99} and therefore we slice across the magnetic field line axis in order to avoid anisotropic effects on the density field. Each 512x512 plane is cut up into 16 images. We repeat this process across all four snapshots, generating $16 \times 512 \times 4 = 32768$ for each of the two simulations. Since the bounds of the simulation box are periodic, the data cube can be displaced in $x$ and $y$ with periodic bounds without injecting sharp edges into the data. In order to present the largest number of uncorrelated images to the network, we shift our box by 64 grid points in $x$ and separately in $y$ before again extracting the 16 images for each consecutive slice along $z$. This technique of ``data enhancement'' is common in neural network training, and yields 32,768 images per cube, or 131,072 per simulation. We perform this procedure on the density cubes for each of the four snapshots for both the super-Alfv\'enic and sub-Alfv\'enic simulations. The data from the first three snapshots comprise the training data, while the data from the last snapshot comprises the test data. We use the last snapshot as the test data because it is a full eddy turnover time from the training snapshots, and thus provides a largely independent test. The image pixel values are scaled by a base 10 logarithm.

In addition to this we make a separate, parallel data set using the same methodology, but with an additional preprocessing step. For each $x-y$ slice of the cube we apply a fast Fourier transform, and set the Fourier power to unity. We then return to the image domain through an inverse fast Fourier transform. Thus, all the images for this fixed Fourier power (FFP) training and test set have no power spectral information.

\begin{figure*}[t!]
\begin{center}
\includegraphics[width=\textwidth]{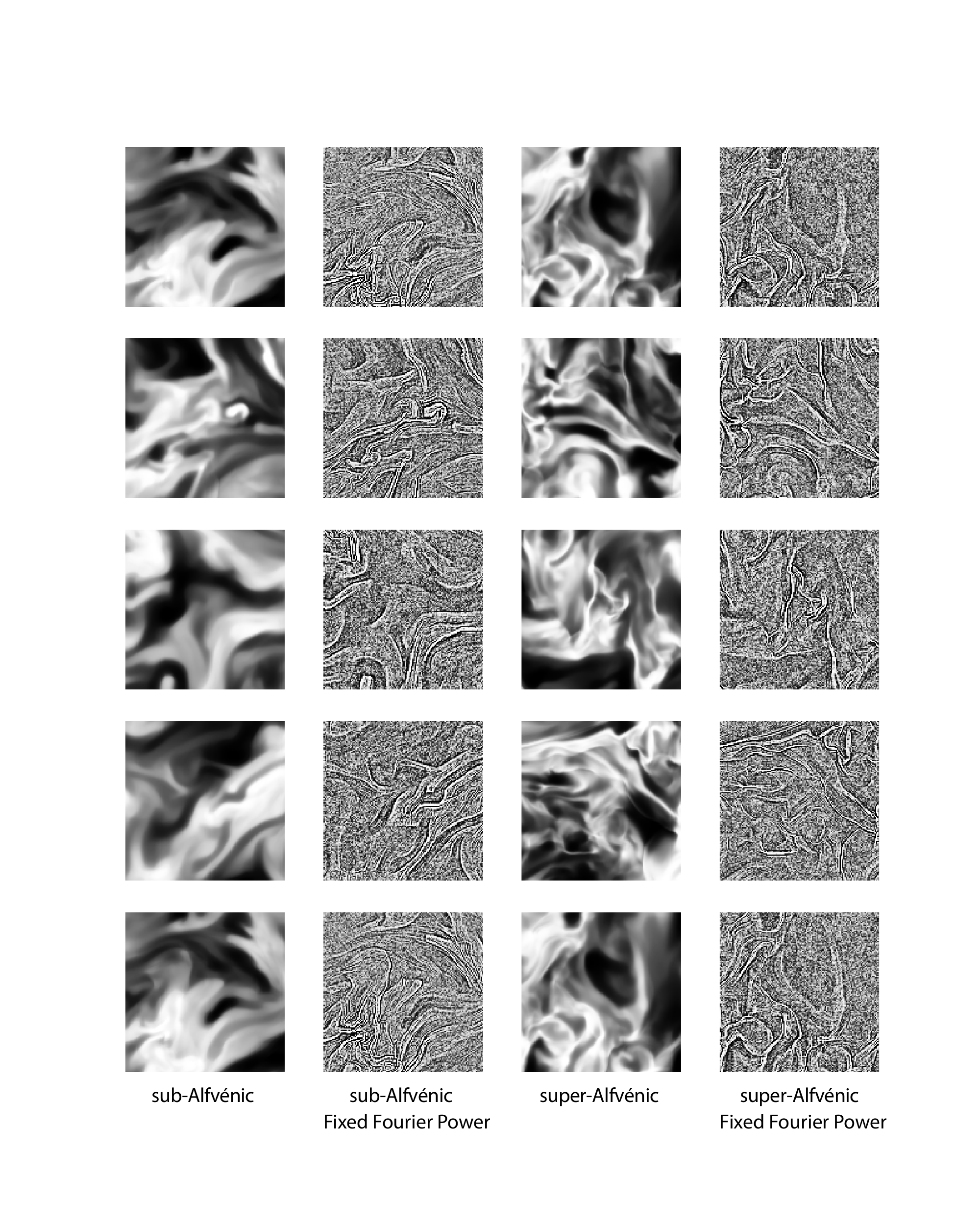}
\caption{Images of example slices. The left two columns are drawn from simulations of sub-Alfv\'enic turbulence, the right two columns from super-Alfv\'enic turbulence. All images are histogram equalized. The 2nd and 4th column have had their power spectra flattened to fixed Fourier power (FFP). The five rows are randomly selected from the data set of 2048 test images.}
\end{center}
\label{images}
\end{figure*}

Lastly, we perform histogram equalization using the exposure method from sci-kit-image \citep{vanderWalt:2014eo}, such that each image has a roughly equal distribution of values from 0 to 1. This is a common preprocessing technique deployed here mainly for the purpose of optimizing the CNN computationally, but it also destroys all information that may be encoded in the pixel amplitude histogram (often called a probability distribution function in the ISM literature, or PDF). For a turbulent density field this PDF should be close to lognormal \citep{Vazquez-Semadeni94a,Padoan97a,Scalo98a,Federrath08a,Burkhart09a}. The procedure is diagramed in Figure \ref{prep}; example images are shown in Figure \ref{images}.

\begin{figure*}[t!]
\begin{center}
\includegraphics[width=\textwidth]{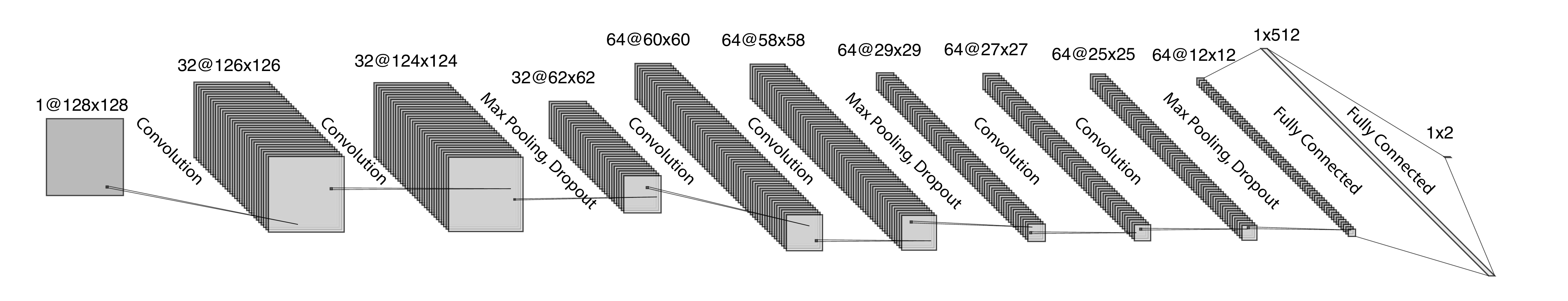}
\caption{The architecture of the neural network used}
\end{center}
\label{netarch}
\end{figure*}

We use a relatively simple CNN, based on an example CNN architecture provided with the Keras package \citep{chollet2015keras} for distinguishing the MNIST set of handwritten digits \citep{LeCun:1998hy}. In this architecture pairs of deep convolutional layers are alternated with pooling layers. 
Dropout, a commonly used method that leaves randomly selected neurons out of the training process, is applied to the pooling layers to reduce overfitting \citep{Srivastava:2014ww}. This sequence is repeated three times until resultant 9,216 neurons can be connected in a dense layer to 512 neurons (see Figure \ref{netarch}). Dropout is also applied to this layer, which leads to the final two classes of neurons, one neuron for sub-Alfv\'enic turbulence, one for super-Alfv\'enic turbulence,. Overall there are 4,858,978 trainable weights in the network. Neuron activation functions were rectified linear units, $f\left(x\right) = max\left(0, x\right)$, except the final layer which was softmax (i.e., a normalized exponential function):
\begin{equation}
    \sigma\left(z\right)_i = \frac{e^{z_i}}{\Sigma_{j=1}^2e^{z_j}}.
\end{equation}

Loss was computed by categorical cross-entropy and adadelta \citep{Zeiler:2012uw} was used for optimization. Adadelta is an optimizer that adjusts learning rates adaptively based on the updates to network gradients. At present it is a typical choice for a network optimizer. The network was trained using batch sizes of 128 images with 12 total epochs (i.e. runs through the full training set).

Deep learning protocols typically reserve a portion of the data as ``validation'' data, and a portion of data as ``test'' data in addition to the training data. The validation data is used to determine how well the network can perform against data it has not seen before. Since a network has many hyperparameters -- how many layers, many optimizer choices, many activation options, etc -- ``test'' data are also set aside to remediate the effects over-fitting through so-called hyperparamter tuning. In this work we did not perform any hyperparamter tuning. The network was modified from the original example network only to make it capable of handling larger 128x128 pixel images. This is because the purpose of this work is not to produce the best possible network for later classification purposes, but rather to determine what information exists, and where it exists, in the images of the simulation available to even a simple network. Our goal is to understand whether and how the CNN is picking up information between two classes of simulations, not to perfect the network. Thus, in this work we have only set aside one set of data, which we call the ``test'' data.

We further explored these data using a method called ``saliency maps'', developed in \cite{Simonyan:2013vv}. Simply put the saliency associated with each pixel in each image is the derivative of the (correct) classification with respect to the pixel amplitude.  The purpose of the method is to determine which parts of each image contributes most to its classification, thus allowing us to build intuition about what features in an image contain the physical underlying information that the network is seizing upon, or whether the network is focused on spurious information. A short exploration confirmed previous work that ``guided backpropagation'' \citep{Springenberg:2014tx} gives the sharpest saliency maps, relatively free of network architecture-induced artifacts. All of the code necessary to extract the images from the simulations, normalize them, build the networks, train the networks, test the networks, and generate figures with data can be found here: \dataset[10.5281/zenodo.3373726]{https://doi.org/10.5281/zenodo.3373726}

\section{Results}
The networks were trained on both the normal and FFP data sets in 15 minutes apiece on a 16 CPU compute node with 60 GB of RAM with an NVIDIA Tesla P100 GPU deployed on the \emph{Google Compute Engine}. The final test accuracy of the network trained on normal images was 97.94\%, and the final test accuracy of the network trained on FFP images was 99.50\%.

While we did not engage in a hyperparameter space search, we did find interesting results regarding the networks and training of the CNN. We found that FFP data took much longer to train, and spent many training epochs with very little decrease in loss function or increase in accuracy on the training data, before rapidly finding a much deeper minimum on the parameter space. Depending on the random seed, we find the network can take many epochs to converge trained on FFP images. We expect this relates to the fact that these images do not have the hierarchy of scales that CNNs are designed to capture. This is to say, the large scale (small wavenumber) power has been removed from these images which makes the CNN less efficient. We urge caution in working with FFP data and CNNs, as we do not yet fully understand what causes the network to converge or fail. We also note that in initial exploratory tests we had degraded the image resolution significantly to speed our analysis, which resulted in significantly worse performance in both normal and FFP analyses. We expect this is due to the fact that small scale features are important to distinguish the super-Alfv\'enic and sub-Alfv\'enic scenarios. Both trained networks, in the model format readable by keras, can be found here: \dataset[10.7910/DVN/UKOPYP]{https://doi.org/10.7910/DVN/UKOPYP}.

\begin{figure*}[t!]
\begin{center}
\includegraphics[width=\textwidth]{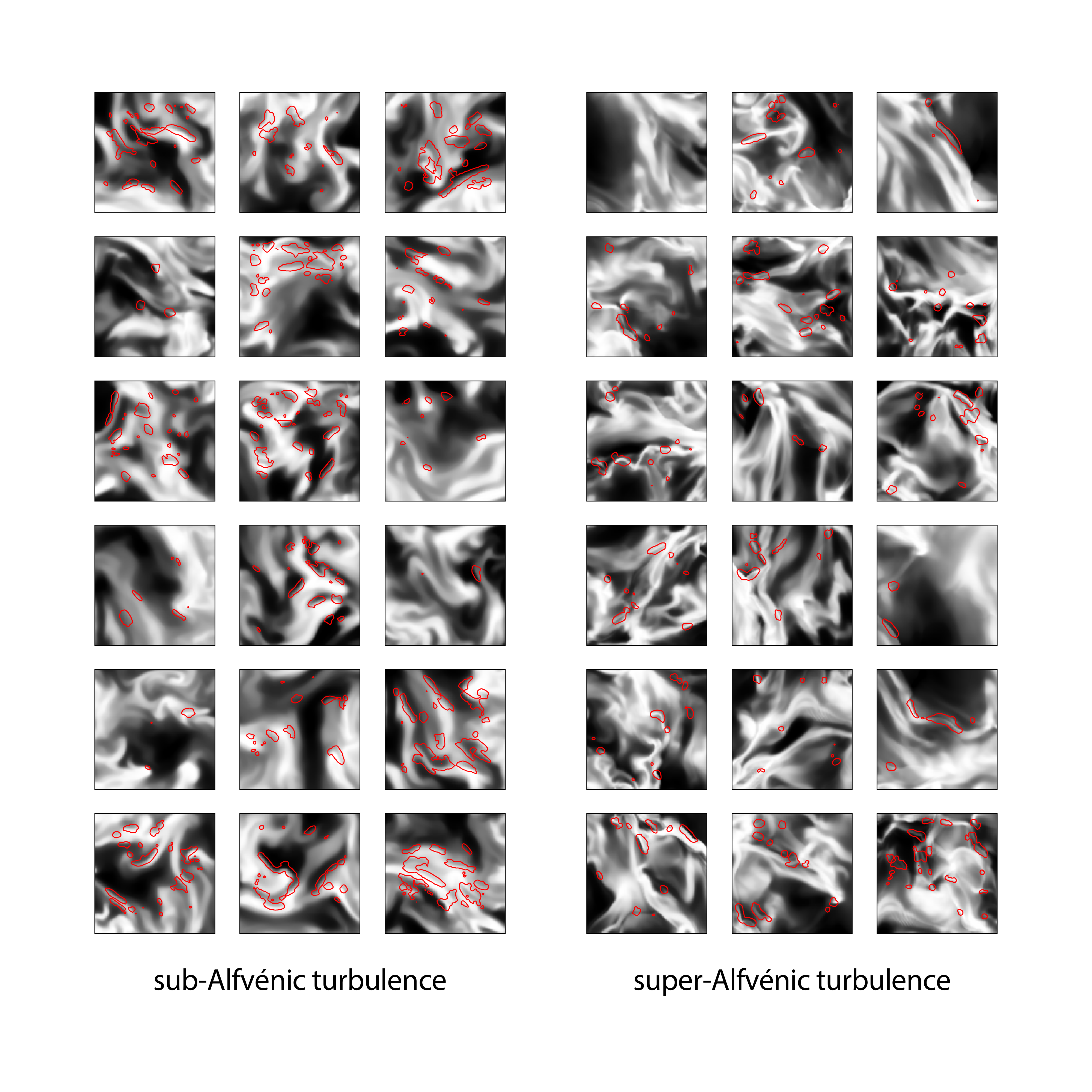}
\caption{Images of 18 randomly selected slices from sub-Alfv\'enic turbulence simulations (left) and 18 from super-Alfv\'enic turbulence simulations (right). A contour of the saliency map for each image at a saliency of 200 is shown in red. On the left we see that the network would find even higher confidence in the sub-Alfv\'enic turbulence label if the dark regions on the boundaries of bright regions were enhanced, thus it is using the smoothness of these boundaries as a marker of sub-Alfv\'enic turbulence. Conversely, on the right, we see the network would be even more confident in the super-Alfv\'enic label if the brightness of sharp, thin density ridges were enhanced. These maps provide a way to understand how the network has learned to label these images, and where the distinguishing morphological information is in the images.}
\end{center}
\label{saliencymaps}
\end{figure*}

An example of 18 randomly selected super-Alfv\'enic and 18 sub-Alfv\'enic sub-images and their associated saliency maps are shown in Figure \ref{saliencymaps}. The saliency map is shown as a single red contour over the image in grayscale. We find that the saliency map very clearly highlights the sharp, ridge-like features in the images in the super-Alfv\'enic case. Conversely, the saliency maps for the sub-Alfv\'enic case highlight dark regions on the edge of bright regions, implying that softening these sharper edges would enhance the sub-Alfv\'enic label confidence. These features highlight how the CNN is distinguishing between the classes of models. 

In the test case we explore here (sub-Alfv\'enic vs. super-Alfv\'enic turbulence), the magnetic field produces fluctuations in the density field  which can be anisotropic in the case of strong sub-Alfv\'enic turbulence and more isotropic for super-Alfv\'enic turbulence.  However because we take slices perpendicular to the mean magnetic field, we are sensitive only to density fluctuations produced by the local magnetic field and not the large scale mean field anisotropy.  The CNN is able to distinguish the local magnetic field fluctuations even after removal of the large scale power and histogram information. This is an unique success for the CNN, as other statistical tools used for magnetic field studies, such as correlation functions, are only sensitive to global magnetic anisotropy \citep{burkhart14}.

\section{Discussion \& Conclusions}

The trained networks were able to tell sub-Alfv\'enic vs. super-Alfv\'enic simulations apart with $\sim98\%$ accuracy when presented only a single 128 $\times$ 128 image, $\sim$0.01\% of the entire simulation volume. These networks were presented with no velocity information, no PDF information, and, in the case of the FFP data sets, no power spectral information. The only information available was embedded in the Fourier phase domain. \textit{This experiment has shown that there is a large amount of valuable information in Fourier phase space not available to techniques that rely on the power spectrum alone.}

These results allow two new avenues for using neural networks. Conservatively, neural networks provide a strong  lower bound on the amount of information about the physical system contained in the image plane. By running a neural network we can determine that there is more information available than is being captured by a given metric if the neural network is easily able to distinguish between the data sets. This can provide encouragement to find more powerful analytic metrics. More speculatively, when trained on much more realistic simulations, one could use these CNNs to determine the underlying physical parameters of the interstellar medium. We caution against taking this approach wholesale, without significant caveats. At present, numerical simulations are  limited in their ability to reproduce observations of complex ISM structures and substructures, in terms of the physics captured, the resolution achievable, the accuracy of synthetic imaging, and our understanding of initial conditions relevant to a given ISM domain. We've shown that in some cases metrics derived from analytics have much less statistical power than CNNs, but they have the distinct advantage of being motivated by theoretical expectations from the physics of the system. Thus it is much easier to quantify in what ways these metrics may be \emph{biased}, or what confounding factors may occur. 

We recommend that CNNs and other deep and machine learning techniques trained directly on pixel-level data be used to measure the observed ISM \emph{in parallel} with analytically based methods like power spectra. If the two methods agree on the underpinning physical parameters like driving scale, sonic mach number and Alfv\'enic mach number, one can have much more confidence in the result. In principle, our method of scraping out information can be extended, such that the image information fed to the CNN is entirely separate from the information in the analytically-based metrics. Such a scenario would give two fully independent measures of the parameters of interest. If these independent metrics gave significantly different answers, this would give a researcher pause in declaring that they had successfully measured a given parameter without determining what systematic uncertainties or biases might be impacting one metric or the other.

Finally, our saliency analysis has shown that neural networks can be very powerful in providing \emph{intuition} for where in the image plane phase information is most distinguishing between different classifications. In the work described here, we found that thin, ridge-like features are the main discriminator the network uses to determine the magnetization of the simulation. In future, these kinds of experiments could allow us to develop better understanding of how different turbulent systems manifest in the image plane. Extrapolating further, saliency can be used with any observational data as well, allowing astronomers to better understand what morphological characteristics (e.g. spiral arm shapes) correlate with what physical characteristics (e.g. star formation rate).

\acknowledgments
B.B. acknowledges the generous support of the Center for Computational Astrophysics in the Flatiron Institute. This work took part under the program The Interstellar Medium Beyond 3D and the program Milky-Way-Gaia of the PSI2 project funded by the IDEX Paris-Saclay, ANR-11-IDEX-0003-02. We acknowledge the Deep Skies Lab as a community of multi-domain experts and collaborators who`ve facilitated an environment of open discussion, idea-generation, and collaboration. This community was important for the development of this project. J. P. thanks Jamie Kinney of Google for help with \emph{Google Compute Engine}. This project made use of Numpy \cite{vanderWalt:2011dp} and Astropy \citep{Astropy-Collaboration13a,2018AJ....156..123A}.
\software{Numpy, astropy}

\end{document}